# Philosophical Solution to P=?NP: P is Equal to NP

Steven Meyer

**smeyer@tdl.com**
March 18, 2016

## Abstract

The P=?NP problem is philosophically solved by showing P is equal to NP in the random access with unit multiply (MRAM) model. It is shown that the MRAM model empirically best models computation hardness. The P=?NP problem is shown to be a scientific rather than a mathematical problem. The assumptions involved in the current definition of the P?=NP problem as a problem involving non deterministic Turing Machines (NDTMs) from axiomatic automata theory are criticized. The problem is also shown to be neither a problem in pure nor applied mathematics. The details of the MRAM model and the well known Hartmanis and Simon construction that shows how to code and simulate NDTMs on MRAM machines is described. Since the computation power of MRAMs is the same as NDTMs, P is equal to NP. The paper shows that the justification for the NDTM P?=NP problem using a letter from Kurt Godel to John Von Neumann is incorrect by showing Von Neumann explicitly rejected automata models of computation hardness and used his computer architecture for modeling computation that is exactly the MRAM model. The paper argues that Deolalikar's scientific solution showing P not equal to NP if assumptions from statistical physics are used, needs to be revisited.

## 1. Introduction

In the early 1970s the informal problem of how to characterize the difficulty of solving problems on digital computers was defined as the "P=?NP" problem. An easy problem could be solved on a deterministic Turing Machine (TM) (Turing[1936]) in a polynomial bounded number of steps. These problems are in the class P. A hard problem requires a non deterministic Turing Machine (NDTM). A NDTM works in parallel trying all computations at once. A problem is then in the class NP if the answer can be checked in polynomial time. The checking is called a decision problem. The P=?NP question asks if problems that seem to require an NDTM can actually be solved in polynomial time on a normal TM.

The P=?NP question was developed by S. Cook (Cook[971], Levin[1973]) who showed that problems in NP could be reduced to the satisfiability problem by a TM in polynomial bounded number of steps. The satisfiability problem takes a set of variables and a collection of clauses and asks if there is a truth assignment that satisfies every clause. The polynomial bound is in the number of variables. The decision problem is easy. Given an assignment for every clause, just check each one. Finding the assignments may be hard requiring exhaustive search. The P=?NP problem gained wide acceptance when R. Karp showed that a number of interesting combinatorial problems could be reduced to satisfiability in polynomial time on a TM (Karp[1971]).

The P=?NP problem has led to progress in axiomatic set theory predicate calculus based complexity by combining many applications of computers into one abstract problem. However, in the process it removed problem specific details and replaced scientific problem solving with axiomatized mathematics. It also does not provide assistance in algorithm design. This paper criticizes the Turing machine model and the methods of formal logic based complexity theory.

It solves the scientific P?=NP problem by showing that P is equal to NP on realistic models of modern computers. It argues the best model of modern computers is the random access memory machine (MRAM) with unit cost multiplication for which NDTMs can be encoded and simulated by the MRAM machine, i.e. realistic models of computation already have the computing power of NDTMs. The paper also defends Deolaliker's scientific solution to the P=?NP problem where he shows P not equal to NP using properties of the statistical physics of Boltzmann gases. Deolaliker's solution and its assumptions and physical observations can be viewed either as a problem in mathematical physics.

This paper should be read in the context of the Popperian school that believes following Niels Bohr that theories are first conceptual then mathematics can be added, but if the mathematics and the conceptual theory differ, the conceptualization has priority. Also, that theory proliferation is positive because it allows theories to be tested by research programme competition (Lakatos[1999]).

## 2. P=?NP Research Program Questionable

P=?NP is not a problem in pure mathematics because the central question is how to characterize and measure computation hardness. Questions of measurement accuracy or suitability are physical problems. The P=?NP research program needs to show that TM based polynomial time is the correct scientific model of computational efficiency. One problem with using automata theory predicate logic is that the philosophy of mathematics in which existence is defined as "can be generated from axioms" is required.

If one uses platonic philosophy of mathematics instead, TM programs do not exist because the halting problem is a Bertrand Russell style paradox. Computable can still mean recursively enumerable because recursive functions exist. If TM programs that do not halt do not exist, then different models of computability are not equivalent and the Church-Turing Thesis becomes meaningless. I am using the argument style of 20th century Swiss mathematician Paul Finsler here (Breger[1992] and Finsler[1996]). It is incorrect to base the empirical question "What makes a computation hard" on one particular philosophy of mathematics.

P=?NP is not a problem in applied mathematics because it does not use mathematics to solve a scientific problem but is itself a theory of calculation. In a sense the P=?NP problem defines computation complexity in order to support the Church-Turing Thesis and the possibility of artificial intelligence (AI). Finsler's Platonism was seen by J. Webb as enough of a problem to criticize it in his book that attempted to show that thought is mechanistic and nothing more than the Church-Turing calculation (Webb[1980], Breger[1992] p. 257). Also J. Lucas review of Webb[1980] in *British Journal for the Philosophy of Science.* (Lucas[1982]).

Peter Naur's criticism of TMs and argument that all non computational thought is being suppressed can be viewed as criticizing the P=?NP research program. See Naur's 2005 Turing Award lecture (Naur[2007]).

## 3. P Equals NP for RAM Machines with Multiply

In the early 1970s there were other models still using the TM predicate calculus automata theory paradigm. The empirically best characterization of computation difficulty is random access machines with unbounded memory cell size and multiplication that only requires one unit of time (called the MRAM model) because it corresponds to operations in physical computers.

The MRAM models were studied by Hartmanis and Simon (Hartmanis[1974] and Hartmanis[1974a]).

The MRAM model is close to modern computer architecture with large random access memories and fast multiplication due to the low level parallelism of multi-issue CPUs with branch prediction. The MRAM model allows unbounded memory cell size and access to bits within memory cells. This allows NDTM configurations to be encoded in memory cells so that MRAMs can simulate NTDMs in Polynomial time (Hartmanis[1974], 38-45).

In the MRAM model P is equal to NP or alternatively the P=?NP problem does not exist in the MRAM model. One might object that MRAM memory cells are unrealistic because they allow storing structures that can be exponential in problem size (the infinity here is finite but unbounded). This type of automata based complexity is abstract. TMs also allow an infinite number of tape squares. Also the Godel style encoding and simulation changes algorithm design from the meta level NDTM world to the world of computer models and application problem "programming".

There are two main reasons the MRAM model is scientifically better than the P=?NP NDTM model. The P=?NP model does not allow human ingenuity in problem specific algorithm design. A trivial example is that sorting can run in linear time by converting a value to a number and storing value k in cell k. This is related to hashing that in the MRAM model does not have the worst case behavior it has in the P=?NP TM model. In the P=?NP automata theory paradigm, ingenuity is used in the transformation and simulation proofs but not in the TM application programs.

Another advantage of the MRAM model is that it corresponds to the von Neumann architecture by allowing low level parallelism without the scheduling and synchronization problems that occur when a large number of simple computers somehow solve a problem in parallel. One can think of the MRAM model as modeling modern CPUs with an unbounded number of instruction issue and coordinate units (sometimes called pipe lines).

See Meyer[2016] for a description of how to utilize modern CPU low level parallelism to implement a Verilog Hardware Description language compiler that is always 30 times faster and sometimes 100 times faster than a compiled simulator that does not utilize the parallelism.

## 4. Von Neumann Rejected Automata Based Measures of Program Hardness

Juri Hartmanis also believed there is a need to justify NDTM based complexity theory in "Godel, von Neumann and the P=?NP Problem" from his April 1989 *The Structural Complexity Column* (Hartmanis[1989]). Hartmanis writes:

There is strong circumstantial evidence that the idea [for] the internally stored program concept, proposed in this report [*Logic Design of a Electronic Computing Instrument* (Aspray[1990], pp. 37-46)] and often attributed to von Neumann was derived from Turing universal machines. Von Neumann has never clarified the origin of these concepts.

Hartmanis' justification is based on a letter from Kurt Godel to John von Neumann in which Godel discusses the problem: "how many Turing machine steps are required to decide if there is a proof of length n for a formula F in predicate calculus". Godel wonders if exhaustive search is required or if only Log N or (log N)**2 steps are needed (p. 6). This section shows that von Neumann explicitly abandoned automata based complexity and was dubious about Godel continuing to hold on to Hilbert's falsified axiomization of everything programme. In fairness, neither Aspray's historical study *John von Neumann and the Origins of Modern Computing.* (Aspray[1990]) nor collections of von Neumann's correspondence were available to Hartmanis in

1989.

## 4.1 Von Neumann as a Philosopher and Mathematical Physicist

I believe there was no interest in using automata theory to study axiomatic logic in the 1950s because the Hilbert programme was viewed as falsified. There was interest especially by von Neumann in the scientific problem of modeling the human nervous system using automata. My argument is based on recent publication of von Neumann's letters and studies analyzing von Neumann's philosophy. Von Neumann saw himself as a philosopher (Neumann[2005], p.16, letter to Fornegura, Dec. 1947). His formalization of quantum mechanics in the late 1920s and early 1930s was undertaken as part of the Hilbert Programme to formalize physics. By the late 1930s von Neumann had given up on the Hilbert Programme and accepted natural philosophy based empiricism of the founders of modern physics. In 1939 von Neumann writes to R. Ortvay (pp. 263-264, also discussed p. 8).

Godel's results mean that there is no "complete" axiomatic system, not even in mathematics, and I believe that there is actually no other consistent interpretation of this complex of questions.

Von Neumann's acceptance of the anti-formalist methods of physics is perhaps best illustrated by a story he would tell in the early 1950s relating Wolfgang Pauli's criticism. "If a mathematical proof is what matters in physics, you would be a great physicist.' (Thirring[2001], p. 5). Von Neumann's change may have occurred from conversations with Nils Bohr at the 1936 Warsaw *New Theories in Physics* Conference (IntCoop[1936]). In a discussion session von Neumann agrees with Bohr that his Hilbert Space formalization is problematic.

In the area of quantum logic, Michael Stoltzner in his article "Opportunistic Axiomatics - von Neumann on the Methodology of Mathematical Physics" (Stoltzner[2001]) argues that von Neumann continued to use axiomatics in his defense of quantum logic into the 1950s disagreeing with my anti-formalism argument in the computing area. However, Stoltzner also argues that von Neumann's opportunism was sometimes philosophically inconsistent, and Stolzner's von Neumann quotations advocating axiomization use examples from quantum mechanics only.

## 4.2 Physicist Opposition to Hilbert's Axiomization of Science

There is a long and well known history of physicist (originally Max Planck and Albert Einstein) opposition to Hilbert's axiomatic physics that von Neumann was familiar with. The dispute started when Planck sent a letter in the early 1890s to Hilbert objecting to Hilbert's axiomization of Kirchhoff's radiation law as being "unsuitable" (Schirrmacher[2010], p. 43). The next dispute arose in in 1895 when Planck hired Ernst Zermelo (later a developer of ZF axiom system) as his assistant. Zermelo proved that reversible physical processes were impossible. The proof is obviously wrong and would not have been popular with a thermodynamics expert such as Planck(Kuhn[1978]. 26-27).

Also, Einstein writes in his 1921 lecture on Geometry criticizing axiomatics:

This view of axioms, advocated by modern axiomatics, purges mathematics of all extraneous elements. ... such an expurgated exposition of mathematics makes it also evident that mathematics as such cannot predicate anything about objects of our intuition or real objects (Einstein[1921]).

Two other areas with opposition to formal axiomatics are first Karl Runge who also taught with Hilbert at Gottingen in the late 19th century and who worked with Planck and thought Hilbert's method of calculating in the continuum was wrong (Schirrmacher[2010]). Second the Bletchley Park code breaking calculations from WW II that directly influenced von Neumann's conception of computing. The Colossus machine was much better than the earlier Turing Enigma machine because William Tutte realized the Turing and Hilbert automata approach of obliviously substituting clauses in a giant array was not as good as low level statistics and iterations (Tutte[1998]). Von Neumann was familiar with Cryptography and listed it as an important application of computers in a funding proposal to Naval Officer Lewis Strauss (Neumann[2005] Oct. 24, 1945 to L. Strauss, p. 237).

## 5. Von Neumann's Characterization of Computation as the MRAM Model

### 5.1 Von Neumann had Moved Beyond Godel's Logic

The previous historical anti-formalist context influenced von Neumann's computer science from his first interest in computation in the early 1940s until his death in 1957. John von Neumann consistently discussed algorithm efficiency as number of operations executed on a digital computer of the von Neumann type. The type of complexity theory asked about in Godel's 1956 letter to von Neumann and justified by Hartmanis using the letter had already been rejected by von Neumann as shown by his writings and discussions.

Even if von Neumann had been healthy, I believe Godel's letter to von Neumann on the problem of "how many Turing Machines steps are required to decide if a proof of length n for a formula F in predicate calculus" (Hartmanis[1989], p. 6) would have fallen on deaf ears because von Neumann's thinking had moved beyond logic models of computation to modern physically realized computers.

Finally, von Neumann had already expressed some skepticism toward Godel's later work. At least they were not working on common projects. In a letter to Institute of Advanced Study director Oswald Veblen recommending Godel for a permanent appointment, von Neumann wrote:

Godel's whole intellectual behavior at present is such, that he may easily do more work in mathematics proper. In fact, I judged, that his probability of doing some is no worse than that of most mathematicians past 35 (Neumann[2005], p. 276, letter to O. Veblen Nov. 30, [????] probably 1945 but maybe 1939).

### 5.2 Von Neumann justification of the MRAM model

In his first draft of his report on EDVAC, von Neumann organized the the computer system as "a central unit to carry out the four basic arithmetic operations (I am following Aspray[1990] here, p.39)." Also "a central control unit to control the proper sequencing of operations and make the individual units work together (p. 39)." Finally, "A memory unit to store both numerical data (...) and numerically coded instructions." Von Neumann was interested in presenting a "logical" description of the stored-program computer rather than an engineering description (p.40). The key to understanding this description as an MRAM model where data describing other algorithms (programs) can be coded in memory cells whose size determined the size of the problem that could be solved (p. 39) is to understand that von Neumann had been concerned and written on various types of infinity. Von Neumann's logical description of a computing device imagined problem size being determined by machine size.

This 19th century type of infinity imagines finite but unbounded machine size determining solvable problem size. Von Neumann assumed his logical machine description would be large enough (unbounded but finite) to solve a given problem. This type of infinity needs to be contrasted with P=?NP type of algebraic infinity where existence and transformation exist because they can be generated from the axioms of set theory.

In a 1946 paper with Herman Goldstine, von Neumann argued that some sort of intuition had to be built into programs instead of using brute force searching (Aspray[1990], p. 62). This needs to be read in contrast to P=?NP where all problem details are removed by a formal automata based method to convert all problems to satisfiability. Von Neumann explicitly rejects automata based complexity in discussing the problem with formal neural networks.

The insight that a formal neuron network can do anything which you can describe in words is a very important insight and simplifies matters enormously at low complication levels. It is by no means certain that it is a simplification on high complication levels. It is perfectly possible that on high complication levels the value of the theorem is in the reverse direction, namely, that you can express logics in terms of these efforts and the converse may not be true. (quoted in Aspray[1990], note 94, p. 321).

Von Neumann argues for using computer programs to analyze properties of Automata in contrast to the P=?NP use of automata to model and explicate computation.

Edward Kohler (Kohler[2000]), p. 118) describes von Neumann's discovery in developing modern computer architecture in an article "Why von Neumann Rejected Carnap's Duality of Information Concepts" as:

Most readers are tempted to regard the claim as trivial that automata can simulate arbitrarily complex behavior, assuming it is described exactly enough. But in fact, describing behavior exactly in the first place constitutes genuine scientific creativity. It is just such a *prima facie* superficial task which von Neumann achieved in his [1945] famous explication of the "von Neumann machine" regarded as the standard architecture for most post World-War-II computers.

## 6. Deolalikar's P!=NP Scientific Solution Merits Further Study

Now that it has been established that P equals NP using the scientifically best model of computation hardness, the Deolalikar scientific solution to the NDTM P=?NP problem becomes much more interesting as a mathematical physics problem and needs to be revisited. Deolalikar showed using a model from statistical physics plus various auxiliary conditions on state space connectivity that P is not equal to NP within the NDTM P=?NP research programme (Deolalikar[2010]). Deolalikar used various theories from disparate areas including statistical physics, state space geometry and properties of Boltzmann gasses to limit the size of the graph from the encoding of satisfiability. It is an empirical question if the various assumptions and conditions on the mathematical objects is justified within the NDTM P=?NP research programme.

Unfortunately, the proof was rejected by the automata theory community, I think, due to two objections by N. Immerman that appeared on R. Lipton's blog (Immerman[2010]). In my view Immerman's criticism is not correct because his two claims of mistakes in Deolalikar's argument are just auxiliary conditions Deolalikar assumes and uses in his scientific solution. The conditions are:

1. Recall that an order on the structure enables the LFP computation (or the Turing machine that runs this computation) to represent tuples in a lexicographical ordering. ... Unfortunately, it is not true that each stage of the fixed point must be order invariant.

2. Second, Deolalikar assumes the nature of computation is limited to simple types of relations (monadic) that limits the number of graph theory edges constructed from the Turing machine model.

The conditions are certainly plausible. Possibly Deolalikar should claim a solution to the NDTM P=?NP problem with a list of assumptions, or maybe the conditions need empirical testing. Another way of stating the disagreement is that in Immerman's view the first problem is an incorrect property of the mapping to a TM and the second monadic relation assumption is "physically" incorrect. In Deolalikar's conception of computation, the conditions correspond to computational reality.

## 7. Conclusion

I would view this philosophical paper a success if it leads to theory proliferation and research programme competition in the computation hardness area. There are interesting TM program speed questions in pure mathematics. The mistake is to assume those problems have any connection to physical reality or empirical questions involving algorithm design. The assumption that a solution to the current NDTM P=?NP problem will change computer science is false. Sometimes expressed as allowing brute force search to be performed rapidly. Another mistake is rejecting proofs of NDTM problems (and other calculation related problems) by excluding or requiring certain assumptions. The rejection of Deolalikar's auxiliary conditions is counter productive and incorrect. A possibly interesting problem in pure mathematics following Finsler's Platonism is to ask what happens to complexity classes if TMs are assumed to not exist.

There are also interesting empirical questions involving mathematical physics. One can imagine an experiment in microphysics that showed statistical models of gases have a specific interaction property that would cause people to view computation differently. Pure mathematics could still study non real gas graph edge connectivity, but it would not be seen as having much interest.

# 8. References

| | |
|---|---|
| Aspray[1990] | Aspray, W. *John von Neumann and The Origins of Modern Computing.* MIT Press, 1990. |
| Breger[1992] | Breger, H. A Restoration that failed: Paul Finsler's theory of sets. In Gillies, D. ed. *Revolutions in Mathematics.* Oxford, 1992, 249-264. |
| Cook[1971] | Cook, S. The Complexity of Theorem-proving Procedures. *STOC'71: Proceedings of the third annual ACM symposium on theory of computing.* ACM Press, 151-158. |
| Deolalikar[2010] | Deolalikar, V. P != NP, HP Research Labs, Palo Alto, Aug. 6, 2010. PDF at URL Jan. 2016: https://www.win.tue.nl/.../Deolalika... |
| Einstein[1921] | Einstein, A. Geometry and Experience. *Lecture before Prussian Academy of Sciences.* Berlin, January 27, 1921, January 2016 URL: www.relativitycalculator.com/pdfs/einstein_geometry_and_experience_1921.pdf |
| Finsler[1996] | Finsler, P. (Booth, D. and Ziegler, R. eds.) *Finsler set theory: Platonism and Circularity.* Birkhauser, 1996. |
| Hartmanis[1974] | Hartmanis, J. and Simon, J. On the Structure of Feasible Computations. *Lecture Notes in Computer Science, Vol. 26.* Also Cornell Ecommons URL Jan. 2016: https://ecommons.cornell.edu/handle/1813/6050, 1974, 1-49. |
| Hartmanis[1974a] | Hartmanis, J. and Simon, J. The Power of Multiplication In Random Access Machines. *15th IEEE Conference on Switching and Automata.* IEEE, Oct. 1974, 13-23. |
| Hartmanis[1989] | Hartmanis, J. Godel, von Neumann and the P=?NP Problem. Periodic *Structural Complexity Column.* April 1989. Cornell Ecommons URL Jan. 2016: eCommons - Cornell digital repository https://ecommons.cornell.edu/handle/1813/6910 |
| Immerman[2010] | Immerman, N. Possible fatal flaws in the finite model part of Deolalikar's proof. From Lipton R. blog, URL Jan. 2016: https://rjlipton.wordpress.com/2010/08/12/fatal-flaws-in-deolalikars-proof |
| IntCoop[1938] | International Institute of Intellectual Co-operation, *New Theories in Physics*, conference proceedings, Warsaw, 1938. |
| Karp[1971] | Karp, R. Reducibility among combinatorial problems. In Miller, R and Thatcher J. (eds.) *Complexity of Computer Programs.* Plenum Press, 1972, 85-103. |
| Kohler[2001] | Kohler, E. Why Von Neumann Rejected Carnap's Duality of Information Concepts. In Redei, M. and Stoltzner, M. (eds.) *John von Neumann and the Foundations of Quantum Physics.* Vienna Circle Institute Yearbook 8, Kluwer, 2001, 97-134. |
| Kuhn[1978] | Kuhn, T. *Black-Body Theory and the Quantum Discontinuity, 1894-1912.* Oxford University Press, 1978. |
| Lakatos[1999] | Lakatos, I. and Feyerabend P. (Motterlini, M. ed) *For and Against Method.* Chicago, 1999. |
| Levin[1973] | Levin. L. Universal sequential Search Problems. *Problems of Information Transmission 9(3).* 1973. |
| Lucas[1982] | Lucas, J. Review of Webb, J. *Mechanism, mentalism, and metamathematics. British Journal for the Philosophy of Science 33,* 441-444. Also Jan. 2016 URL: http://users.ox.ac.uk/~rlucas/Godel/webb.html |
| Meyer[2016] | Meyer, S. CVC Verilog Compiler - Fast Complex Language Compilers can be Simple. Unpublished, Jan 2016 URL: |

www.tdl.com/~smeyer/docs/cvc-ver-comp.pdf, 2016. '

Naur[2007]          Naur, P. Computing versus human thinking. *comm. ACM 50, 1(2007)*, 85-94.

Neumann[2005]       Von Neumann, J. (Redei, M. ed.) *John Von Neumann: Selected Letters.* History of Mathematics Series, Vol. 27, American Mathematical Society, 2005.

Schirrmacher[2010]  Schirrmacher, A. Theoretiker zwischen mathematischer und experimenteller Physik. *Max Planck und die Moderne Physik.* (Hoffman, D. ed), Springer, 2010, 35-48.

Stoltzner[2001]     Stoltzner, M. Opportunistic Axiomatics - von Neumann on the Methodology of Mathematical Physics. In Redei, M. and Stoltzner, M. (eds.) *John von Neumann and the Foundations of Quantum Physics.* Vienna Circle Institute Yearbook 8, Kluwer, 2001, 5-10.

Thirring[2001]      Thirring, W. J. v. Neumann's Influence in Mathematical Physics. In Redei, M. and Stoltzner, M. (eds.) *John von Neumann and the Foundations of Quantum Physics.* Vienna Circle Institute Yearbook 8, Kluwer, 2001, 5-10.

Turing[1936]        Turing, A. On computable numbers, with an application to the Entscheidungsproblem. *Proceedings of the London Mathematical Society, Series 2, 42(1936-7),* 230-265.

Tutte[1998]         Tutte, W. *Fish and I.* Known as the Fish Lecture on the Bletchley Park Colossus machine, 1998, URL Jan 2016: https://en.wikipedia.org/wiki/W._T._Tutte, 9th entry in Sources.

Webb[1990]          Webb, J. *Mechanism, mentalism, and metamathematics.* Reidel, 1980.